\documentclass[conference]{IEEEtran}
\IEEEoverridecommandlockouts
\usepackage{cite}
\usepackage{amsmath,amssymb,amsfonts}
\usepackage{algorithmic}
\usepackage[ruled, linesnumbered]{algorithm2e}
\usepackage{graphicx}
\usepackage{textcomp}
\usepackage{subfigure} 
\newtheorem{definition}{Definition}
\usepackage{xcolor}
\def\BibTeX{{\rm B\kern-.05em{\sc i\kern-.025em b}\kern-.08em
		T\kern-.1667em\lower.7ex\hbox{E}\kern-.125emX}}
\begin{document}
	
\title{\huge An Improved Virtual Force Approach for UAV Deployment and Resource Allocation in Emergency Communications\\}
\author{
\IEEEauthorblockN{Hongying Guo, Li Wang,~\IEEEmembership{Senior Member,~IEEE}, Ruoguang Li, \IEEEmembership{Member, IEEE}, \\Luyang Hou, Lianming Xu, and Aiguo Fei\\}
\thanks{
Hongying Guo, Li Wang, Luyang Hou, and Aiguo Fei are with School of Computer Science (National Pilot Software Engineering School), Beijing University of Posts and Telecommunications, Beijing 100876, China (e-mail: 
\{guohongying, liwang, luyang.hou, aiguofei\}@bupt.edu.cn). \emph{(Corresponding author: Li Wang.)}

Ruoguang Li is with the National Mobile Communications Research Laboratory, Southeast University, Nanjing 210096, China (e-mail: ruoguangli@seu.edu.cn).

Lianming Xu is with School of Electronic Engineering, Beijing University of Posts and Telecommunications, Beijing 100876, China (e-mail: xulianming@bupt.edu.cn).
}
\vspace{-5mm}
}

\maketitle
\begin{abstract}
In this paper, we consider an unmanned aerial vehicle (UAV)-enabled emergency communication system, which establishes temporary communication link with users equipment (UEs) in a typical disaster environment with mountainous forest and obstacles. Towards this end, a joint deployment, power allocation, and user association optimization problem is formulated to maximize the total transmission rate, while considering the demand of each UE and the disaster environment characteristics. Then, an alternating optimization algorithm is proposed by integrating coalition game and virtual force approach which captures the impact of the demand priority of UEs and the obstacles to the flight path and consumed power. Simulation results demonstrate that the computation time consumed by our proposed algorithm is only $5.6\%$ of the traditional heuristic algorithms, which validates its effectiveness in disaster scenarios.
\end{abstract}


\section{Introduction}

Unmanned Aerial Vehicles (UAVs), which can provide wireless coverage from sky for Users Equipment (UEs), are expected to be a promising technology of the six generation (6G) mobile networks. Due to the inherent advantanges in terms of the mobility and flexibility, UAV offers real-time and cost-efficient services for numerous applications, especially accomodating the demand of emergency communication where the terrestrial infrastructures are desptroyed during the natural disasters\cite{tian2023uav123}.  In emergency situations, the rapid deployment of UAVs to achieve communication coverage holds great significance. However, due to the limited availability of communication resources, it is crucial to schedule and utilize these resources effectively\cite{wang2020edge}. In previous UAV deployment optimization studies, Wang et al. proposed a multiple UAVs path planing approach based on particle swarm optimization (PSO) algorithm to shorten the path distance for fast data collection\cite{9771595}. Ghamry et al. utilized PSO algorithm for optimal paths in forest fire fighting missions\cite{7991527}. Liu et al. employed a genetic algorithm (GA) for efficient deployment of the minimum UAVs to maximize wireless coverage in emergency rescue scenarios\cite{8720417}. Luan et al. introduced the virtual force field to optimize UAV deployment, considering the trade-off between UAV load balancing and transmission rate\cite{luan2021joint}.

Regarding communication resource allocation optimization, Wang et al. aimed to minimize the overall transmission cost while guaranteeing users' requirement by proposing a hypergraph-based local search algorithm\cite{wang2016hypergraph}. Li et al. proposed a method to optimize resource allocation in UAV-aided communication networks by employing flexible power control and bandwidth allocation\cite{10226040}. Additionally, various studies have investigated joint optimization of UAV deployment and communication resource allocation. Wu et al. presented a joint optimization of UAVs deployment and power allocation in the multi-level quality of service driven scheme\cite{9685172}. Li et al. proposed a heuristic algorithm based on matching theory in cellular networks, aiming to jointly optimize resource allocation and UAV trajectory for maximizing total energy efficiency in downlink wireless networks with quality of service requirements\cite{9322351}.

While the previous studies have made efforts to tackle UAV deployment and resource allocation problems using heuristic algorithms, their high computational complexity poses a challenge, particularly in time-sensitive scenarios like emergency communications. Furthermore, the presence of obstacles in the rescue area cannot be overlooked.
In this paper, we consider a UAV-enabled emergency communication system in a disaster environment with mountainous forest and obstacles. Our objective is to maximize the total transmission rate. To achieve this, we propose an alternating optimization algorithm that integrates the coalition game and virtual force approach. This algorithm considers the presence of obstacles, incorporates individual preferences in user association and jointly optimizes the deployment, power allocation, and user association. Simulation results  show that the proposed method significantly reduces computation time compared to the conventional heuristic algorithm, making it highly effective in disaster scenarios.

\section{System Model and Problem Formulation}
\subsection{System Model}
\par 
We consider a multi-UAV-aided cooperative network for emergency communication in a mountainous forest area with various obstacles. The network comprises $N$ UAVs forming a connected airborne network, aiming to provide wireless communication services to $M$ UEs affected by the emergency situation.
Specifically, in the multi-UAV-aided cooperative network, UAVs establish communication links with their respective connected UEs by utilizing different frequency bands. The set of UAVs, UEs, and obstacles are denoted by $\mathcal{I}$,  $\mathcal{K}$, and $\mathcal{O}$, respectively. The disaster area is assumed to have a rectangular shape $(\Lambda_x \times \Lambda_y)$. We consider a two-dimensional
(2D) Cartesian coordinate system, where UE $k$ is located at $\mathbf{w}_k=(x_k,y_k),\forall k \in \mathcal{K}$, $0\leq x_k\leq \Lambda_x$ and $0\leq y_k \leq \Lambda_y$. 
And UAV $i$ is located at $\mathbf{z}_i=(x_i,y_i),\forall i \in \mathcal{I}$, where $0\leq x_i\leq \Lambda_x$ and $0\leq y_i \leq \Lambda_y$. The coordinate set of UAV is denoted by ${\mathcal{Z}}$. Besides, $\mathcal{P}$ denotes the UAV transmit power set, in which $p_{i,k}$ is the transmit power from UAV $i$ to UE $k$. 
Furthermore, we introduce a matrix $\mathcal{C}=[c_{i,k}]_{N\times M}$, and
the binary variable $c_{i, k}$ characterizes the association between the UAV $i$ and UE $k$, i.e., $c_{i, k}=1$ if UE $k$ is associated with UAV $i$, otherwise  $c_{i, k}=0$. We assume that UAVs employ the frequency division multiple access (FDMA) technique, where each UAV $i$ occupies a total bandwidth of $B_i$. This bandwidth is equally divided among the UEs associated with UAV $i$, resulting in the bandwidth allocated to UE $k$ as $b_k=B_i/\sum_{k\in\mathcal{K}}c_{i,k}$.
\vspace{1mm}
\par Due to the existence of vegetation in the mountainous forest area, the transmission path loss $L_{i,k}$ between UAV $i$ and UE $k$ can be expressed by\cite{6484996}
\setlength\abovedisplayskip{0.2cm}
\setlength\belowdisplayskip{0.2cm}
\begin{equation}
	\label{sec_system_model}
	L_{i, k}=L_{i, k}^{F S P L}+10 \alpha \log \left(\frac{d_{i, k}}{d_{0}}\right)+X_{\sigma}+L_{i, k}^{\text {Slant}},
\end{equation}
where $X_{\sigma}\sim N \left(0,\sigma_x^2\right) $ is a zero-mean Gaussian random variable with variance $\sigma^2_x$, $d_{0}$ is the reference distance, $d_{i,k}$ is the distance between UAV $i$ and UE $k$, and $\alpha$ is the path loss exponent. $L_{i, k}^{F S P L}$ describes the free space propagation loss between UAV $i$ and UE $k$. $L_{i,k}^{\text {Slant}}$ is the excess loss in the forest scenario based on ITU-R Recommendation P.833-9\cite{RN07}. Specifically, $L_{i, k}^{F S P L}$ and $L_{i,k}^{\text {Slant}}$ can be expressed by $L_{i, k}^{F S P L}=20 \log \left(\frac{4 \pi f d_{i,k}}{c}\right)$ and $L_{i, k}^{\text {Slant }}=A f^{C} {d_{i, k}^{D}}(\theta_{i,k}+E)^{G}$, where $c$ is the speed of light, $\theta_{i,k}$ is the radio path elevation angle, and $f$ is the carrier frequency. $A$, $C$, $D$, $E$, and $G$ are empirical parameters, respectively. The channel between UAV $i$ and UE $k$ is given by $h_{i,k}=\sqrt{L_{i,k}}g_{i,k}$, where $g_{i,k}$ is the small-scale fading gain. The received SNR from UAV $i$ to UE $k$ is expressed by
	\begin{equation}
	\gamma_{i, k}=\frac{p_{i, k}\left|h_{i, k}\right|^2}{\sigma^2},
	\label{SINR}
\end{equation}
where $\sigma^2$ denotes the variance of additive white Gaussian noise (AWGN). Therefore, the achievable rate between the UAV $i$ and UE $k$ can be written as:
		\begin{equation}
	R_{i,k}=b_k\log_2\left(1+\gamma_{i,k}\right).
	\label{Rate}
\end{equation}
\vspace{-10mm}
\subsection{Problem Formulation}
\par To ensure communication service for all on-ground UEs, we formulate an optimization problem that addresses UAV deployment, power allocation, and user association. The objective is to maximize the total transmission rate while considering constraints such as received SNR and transmit power limits. Mathematically, the optimization problem can be stated as follows:

\begin{small}
\begin{subequations}\small
	\label{optimization_label}
	\begin{align} 
		&\max _{{\mathcal{Z}}, \mathcal{P},\mathcal{C}} \sum_{i \in \mathcal{I}} \sum_{k \in \mathcal{K}} R_{i,k} \\
		\text{s.t.}\quad
		&\sum_{i=1}^N c_{i, k} \leq 1, \forall k \in \mathcal{K}  \label{opt-c2},\\
		&\sum_{k=1}^M c_{i,k}p_{i,k} \leq P_i, \forall i \in \mathcal{I},\label{opt-power}\\
		& \gamma_{k}^{th} \leq c_{i, k}\gamma_{i,k}, \forall k \in \mathcal{K}, \exists i \in \mathcal{I},\label{opt-link}\\
		&c_{i,k} \in\{0,1\}, \forall i \in \mathcal{I},\forall k \in \mathcal{K},\label{opt-c1}\\
		&x_i \in\left[0, \Lambda_x\right], y_i \in\left[0, \Lambda_y\right],\forall i \in \mathcal{I}\label{opt-range},
	\end{align}
\end{subequations}
\end{small}where \eqref{opt-c2} ensure that each UE is served by at most one UAV. \eqref{opt-power} are the transmit power limit of UAV $i$, where $P_i$ is the maximum transmit power of UAV $i$. \eqref{opt-link} are the SNR requirement for UE $k$. \eqref{opt-c1} are the Boolean constraints for UAV-UE association and constraints \eqref{opt-range} guarantee that UAV $i$ is deployed in the prescribed area. 
\vspace{-3mm}
\section{Proposed Solution}
The optimization problem \eqref{optimization_label} is a non-convex program, which is NP-hard and cannot be efficiently solved by standard convex optimization tools. To address this optimization problem, in this section, we first initialize the UAV deployment using K-Means algorithm, and then iteratively
solve the problem \eqref{optimization_label} by utilizing coaltion game and virtual force approach.  
\vspace{-5mm}
\subsection{K-Means for UAV Deployment Initialization}
\par First, we employ the K-Means algorithm to determine the required number of UAVs for covering all UEs \cite{8796414}. Assume that the transmit power is equally allocated among different UAVs, and this power allocation is denoted as $\mathcal{P}^{\text{ini}}$. We start by setting the number of clusters to $N=N_0$. We randomly select $N_0$ UEs from $\mathcal{K}$ as cluster centroids, which represent the initial locations of the UAVs. UEs are assigned to the nearest cluster centroid. Centroids are updated by averaging the locations of clustered UEs until convergence is reached. We then check whether the current deployment fulfills the communication demands of UEs. If all UEs' communication demands are met, we obtain the initial deployment $\mathcal{Z}^{\text{ini}}$ that serves the UEs while utilizing the minimum number of UAVs. However, if the communication demands are not satisfied for all UEs, we increase the number of UAVs ($N_0 \leftarrow N_0+1$) and repeat the process until the communication demands of all UEs are successfully fulfilled.
\vspace{-3mm}
\subsection{Coalition Game for User Association}
\label{sec_game}
The user association matrix $\mathcal{C}^{}$ is determined by maximizing total transmission rate with coalition game\cite{sami2019user}. The corresponding optimization problem is expressed as
\setlength\abovedisplayskip{-0.cm}
\setlength\belowdisplayskip{-0.cm}
	\begin{subequations}
		\begin{align}
			&\max _{{\mathcal{C}}} \sum_{i \in \mathcal{I}} \sum_{k \in \mathcal{K}} c_{i,k}R_{i,k}\\
			&\text{s.t.}\quad \eqref{opt-c2},\eqref{opt-power},\eqref{opt-link},\eqref{opt-c1}.
		\end{align}
	\end{subequations}
The total transmission rate increases when certain UEs transfer from heavily-loaded coalitions to lightly-loaded coalitions. This is because associating more UEs with a single UAV reduces the available power of each UE. When transferring the association of UE $k$ from UAV ${i'}$ to UAV ${i}$, the following conditions must be satisfied:
	\begin{equation}
		\label{transfer_factor}
		\sum_{i \in \mathcal{I}}{c_{i,k}R_{i,k}} - \sum_{i' \in \mathcal{I}}{c_{i',k}R_{i',k}}>0, i,i' \in \mathcal{I}, i \neq i'
		.
	\end{equation}  
The condition guarantees that the transfer increases the total transmission rate of the network.
\begin{definition}
	\label{game}
	(Coalition Game). The coalition $s_{j}$ is defined as the set of UEs associated with UAV $j$. Define the coalition game by $(S,U,\mathcal{I},\mathcal{K})$, where $S=\{s_1,\dots,s_i,\dots,s_N\}$ is the coalition set and ${U}$ is the set of utilities of each transfer, the utility $u_{k,i}$ of UE $k$ transfer from coalition $s_{i'}$ to coalition $s_{i}$  is defined by:
		\begin{equation}
			\label{pre}
			u_{k,i} =\sum_{i \in \mathcal{I}}{c_{i,k}R_{i,k}} - \sum_{i' \in \mathcal{I}}{c_{i',k}R_{i',k}}, i,i' \in \mathcal{I}, i \neq i'.
		\end{equation}
\end{definition}And UAVs can build preference relation based on the estimated utility\cite{sami2019user}.
\begin{definition}
	\label{preference}
	(Preference relation). UAV $i$ ranks the UEs based on a  preference relation denoted as $\succ$. The notation $k \succ_{i} l$ indicates that UAV $i$ prefers UE $k$ over UE $l$, which can be interpreted as the utility  provided by associating with UE $k$ being higher than that provided by associating with UE $l$, i.e., $
	u_{k,i}>u_{l,i}$.
\end{definition}

Based on Definition \ref{game} and \ref{preference}, we can obtain the optimal user association through the following steps: first, we set $t=0$ and initialize the coalition as $S^{t}$, where each UE is associated with the UAV that can provide the largest received SNR. Next, we check the transfer condition to identify UEs that could potentially transfer to another coalition based on \eqref{transfer_factor}. Each UAV selects the UE with preference relation defined in Definition \eqref{preference}, resulting in a new coalition partition $S^{(t+1)}$. We repeat this process until the increment of the total utility $U$ is below a predefined threshold $\varepsilon$. The details of the process are summarized in Algorithm \ref{algorithm_UA}.
\begin{small}
\begin{algorithm}[t]\small
	\label{algorithm_UA}
	\textbf{Initialization:} Initiate $S^{(t)}=\{s_1^{(t)},\dots,s_N^{(t)}\},t=0$. \\
	Calculate $u_{k,i},\forall k \in \mathcal{K},\forall i \in \mathcal{I}$ based on \eqref{pre}, obtain $U^{(t)}$.\\
	\textbf{Coalition Game:}\\
	\Repeat{$U^{(t)}-U^{(t-1)}<\varepsilon$}{
		identify the UEs that satisfy the transfer condition based on \eqref{transfer_factor},\\
		each UAV selects the UE with the largest estimated utility based on Definition \eqref{preference},\\
		update $s_{n}^{(t+1)}$, $S^{(t+1)} \leftarrow \{s_{1}^{(t+1)},\dots,s_n^{(t+1)},\dots,s_{N}^{(t+1)}\}$,\\
		update $U^{(t+1)}$ based on \eqref{pre},\\
		$t \leftarrow t+1$.\\
	}
	$S^{*}=S^{(t)}$, update $c_{n,k},\forall n \in \mathcal{N},\forall k \in \mathcal{K}$, obtain $\mathcal{C}$.\\
	output $\mathcal{C}^{}$.
	\caption{Coalition Game for User Association}
\end{algorithm}
\end{small}
\vspace{-3mm}
\subsection{Virtual Force Approach for UAV Deployment and Power Allocation}

%\subsection{Virtual Force Model}
\setlength{\abovecaptionskip}{-0.cm}
\setlength{\belowcaptionskip}{-0.5cm}
\begin{figure}[!hbt] 
	\centering
	\subfigure[Attractive forces]
	{	
		\begin{minipage}[!hbt]{0.4\linewidth}
			\raggedleft
			\label{fig_attractive_force}
			\includegraphics[width=1.3in]{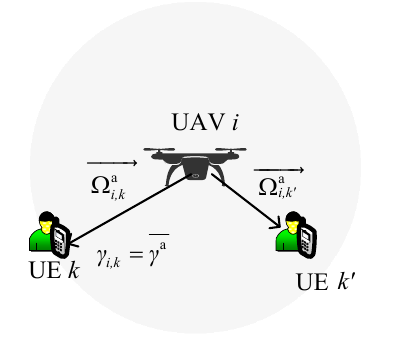}
		\end{minipage}
	}
	\subfigure[Repulsive forces]
	{
		\begin{minipage}[!hbt]{0.4\linewidth}
			\raggedright
			\label{fig_repulsive_force}
			\includegraphics[width=1.3in]{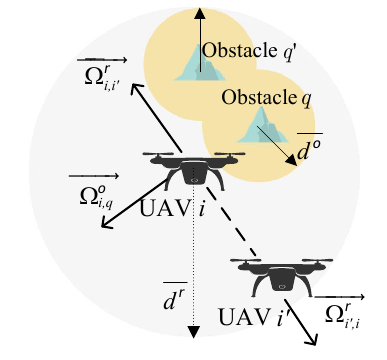}	
		\end{minipage}
	}
	\centering
	\caption{Illustration of virtual forces.}
	\label{fig_force}
	\vspace{-5mm}
\end{figure}
After the user association solution is updated, the UAV deployment and power allocation are jointly optimized by the virtual force approach. Specifically, as shown in Fig. \ref{fig_force}, there are three kinds of virtual forces considered in our model: (1) the attractive force towards UEs $\overrightarrow{{\Omega}^{{\mathrm{a}}}_{i,k}}$; (2) the repulsive force between UAVs $\overrightarrow{\Omega^{\mathrm{r}}_{i,j}}$; and (3) the repulsive force to avoid obstacles $\overrightarrow{\Omega^{\mathrm{o}}_{i,q}}$.
\subsubsection{The Attractive Force Towards UEs} As illustrated in Fig.  \ref{fig_force}\subref{fig_attractive_force}, the attractive force drags the UAV towards UE to provide communication services, which characterizes the on-demand coverage and transmit power requirement of each UE. The attractive force of UAV $i$ to UE $k$ is formulated by the universal gravitation model, represented by
\begin{small}
	\begin{equation}
		\label{fa}
		\begin{split}
			&\overrightarrow{{\Omega}^{\mathrm{a}}_{i,k}}= 
			\begin{cases}
				\frac{\overrightarrow{\lambda_{\mathrm{a}}\cdot p_{i, k} \cdot\left(\gamma_{i,k}-\gamma_{k}^{th}\right)}}{d_{i,k}^2}, \gamma_{i,k}>\bar{\gamma}^{\mathrm{a}} \text { and }\sum\limits_{k' \in \mathcal{K}} c_{i,k'}p_{i,k'} \leq P_i,\\
				0,\text{else},
			\end{cases}
		\end{split}
	\end{equation}
\end{small}where ${\lambda}_{\mathrm{a}}$ is the attractive force factor, $d_{i,k}$ is the distance between UAV $i$ and UE $k$. $\bar{\gamma}^{\mathrm{a}}$ is the received SNR threshold for UE $k$ to attract UAV $i$. It is worth noting that when the $\gamma_{i,k}$ exceeds $\bar{\gamma}^{\mathrm{a}}$ and the actual transmit power of UAV $i$ is less than the maximum transmit power $P_i$, the attractive force $\overrightarrow{\Omega^{\mathrm{a}}_{i,k}}$ works.

\subsubsection{The Repulsive Force Between UAVs}  Fig.  \ref{fig_force}\subref{fig_repulsive_force} represents the repulsive force between different UAVs, preventing UAVs from colliding with each other and maintaining the safety distance of two UAVs with a predefined value. Based on Hooke’s Law\cite{8796414}, we model the repulsive force the between UAV $i$ and $i'$ ($i\not = i'$) as
\begin{small}
	\begin{equation}
		\label{fr}
		\begin{split}
			&\overrightarrow{\Omega^{\mathrm{r}}_{i,i'}}= 
			\begin{cases}
				\overrightarrow{\lambda_{\mathrm{r}}\left(\bar{d}^{\mathrm{r}}-d_{i,i'}\right)}, d_{i,i'}<\bar{d}^{\mathrm{r}} \text { and } \sum\limits_{k \in \mathcal{K}} c_{i,k}p_{i,k} \leq P_i, \\
				0, \text { else },	
			\end{cases}
		\end{split}
	\end{equation}
\end{small}where $\lambda_{\mathrm{r}}$ is the repulsive force factor, $d_{i,i'}$ is the distance between the UAV $i$ and UAV $i'$. When $d_{i,i'}$ is less than a predefined range $\bar{d}^{\mathrm{r}}$ and the actual transmit power of UAV $i$ is less than the maximum transmit power $P_i$, the repulsive force $\overrightarrow{\Omega^{\mathrm{r}}_{i,i'}}$ works and pushes UAV $i$ away from UAV $i'$.
\subsubsection{The Repulsive Force to Avoid Obstacle} As shown in Fig. \ref{fig_force}\subref{fig_repulsive_force}, if there are obstacles in front of UAV $i$, the repulsive force works, rendering UAV $i$ move to an opposite direction to avoid obstacles. Specifically, the repulsive force between UAV $i$ and obstacle $q$ is modeled as
\begin{small}
	\begin{equation}
		\label{fo}
		\begin{split}
			&\overrightarrow{\Omega^{\mathrm{o}}_{i,q}}=
			\begin{cases}
				\overrightarrow{\lambda_{\mathrm{r}} \left(\bar{d}^{\mathrm{o}}-d_{i,q}\right)}, d_{i,q}<\bar{d}^{\mathrm{o}} \text { and } \sum\limits_{k \in \mathcal{K}} c_{i,k}p_{i,k} \leq P_i,\\
				0, \text { else },	
			\end{cases}
		\end{split}
	\end{equation}
\end{small}where $\bar{d}^{\mathrm{o}}$ is a predefined safe distance to the obstacle and $d_{i,q}$ is the distance between UAV $i$ and the obstacle. Furthermore, we treat the boundary of the prescribed area as the obstacle to ensure that the UAV does not move out of the area.
\par Based on \eqref{fa}, \eqref{fr}, and \eqref{fo}, the aggregated force of the UAV $i$ is calculated by,
	\begin{equation}
		\label{fsum}
		\overrightarrow{\Omega_i(\boldsymbol{z}_i, \{p_{i,k}\})}=\sum_{k \in \mathcal{K}}\overrightarrow{\Omega^{\mathrm{a}}_{i,k}}+\sum_{i'\in\mathcal{N}, i'\not= i}\overrightarrow{\Omega^{\mathrm{r}}_{i,i'}}+\sum_{q \in \mathcal{Q}}{\overrightarrow{\Omega^{\mathrm{o}}_{i,q}}}.
	\end{equation}
It can be observed that \eqref{fsum} is a function with respect to the UAV $i$'s location and the corresponding transmit power subject to the constraints in \eqref{optimization_label}. 
According to Newton’s second law of motion, the change of velocity in the time duration $\Delta t$ is governed by the resultant force $\overrightarrow{\Omega_i(\boldsymbol{z}_i, \{p_{i,k}\})}$ acting on UAV $i$\cite{8796414}. We assume that the virtual mass of UAV $i$ is $m_i$, which is normalized as 1. Then, we have
\begin{equation}
	\begin{split}
		\label{v1}
		\overrightarrow{\Delta v_i(\boldsymbol{z}_i, \{p_{i,k}\})}=\overrightarrow{\Omega_i(\boldsymbol{z}_i, \{p_{i,k}\})} \cdot \frac{\Delta t}{m_i}=\overrightarrow{\Omega_i(\boldsymbol{z}_i, \{p_{i,k}\})} \cdot \Delta t.
	\end{split}
\end{equation}
We assume that the velocity is controlled periodically, e.g., $\Delta t=1$ second. 
According to \eqref{v1}, $\overrightarrow{\Delta v_i(\boldsymbol{z}_i, \{p_{i,k}\})}$ is positively correlated with $\overrightarrow{\Omega_i(\boldsymbol{z}_i, \{p_{i,k}\})}$ and it may have infinite magnitude which violates the dynamics of UAVs. 
Thus, it is necessary to map $\overrightarrow{\Delta v_i(\boldsymbol{z}_i, \{p_{i,k}\})}$ to a finite range. To perform this mapping, we select the trigonometric function $\rm arctan( )$ \cite{8796414}. The velocity is given by
	\begin{equation}
		\label{v2}
		\overrightarrow{{v_i}(\boldsymbol{z}_i, \{p_{i,k}\})}=\arctan (\overrightarrow{\Delta v_i(\boldsymbol{z}_i, \{p_{i,k}\})})\cdot\frac{2v_{max}}{\pi},
	\end{equation}
where $v_{max}$ is the maximum velocity achieved by UAV $i$. The velcoity $\overrightarrow{{v_i}(\boldsymbol{z}_i, \{p_{i,k}\})}$ includes all the constraints in $\overrightarrow{\Omega_i(\boldsymbol{z}_i, \{p_{i,k}\})}$, which pushes the UAV $i$ towards UEs and simultaneously optimizes the power allocation to increase the received SNR as much as possible, resulting in an improved total transmission rate. Therefore, the purpose of virtual force is to maximize the transmission rate in \eqref{optimization_label} by equivalently updating  $\overrightarrow{{v_i}(\boldsymbol{z}_i, \{p_{i,k}\})}$. When the variance of $\overrightarrow{{v_i}(\boldsymbol{z}_i, \{p_{i,k}\})}$ is small, the UAV arrives at the optimal location and power allocation, and the maximum transmission rate is obtained.
\begin{figure*}[!htb]
	\centering
	\subfigure[5 UAVs, coverage rate = $82\%$,   
	transmission rate = 10.72 Mbps.]
	{
		\begin{minipage}[b]{0.3\linewidth}
			\centering
			\includegraphics[width=2.in,height=1.7in]{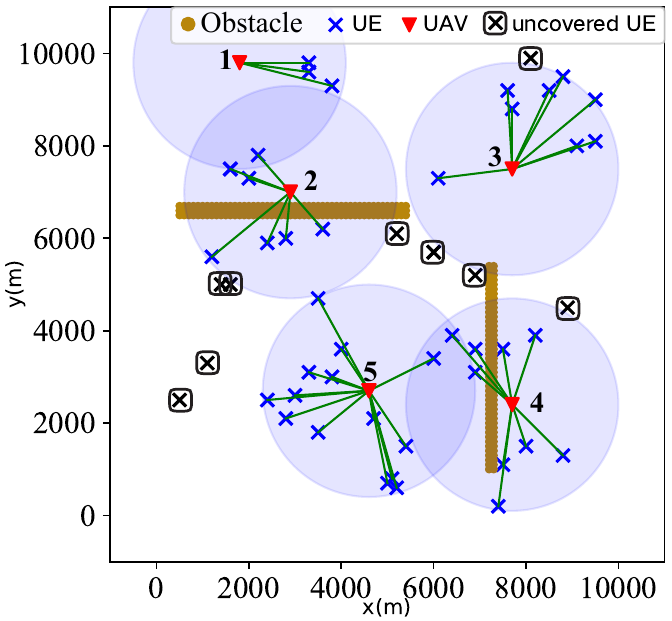}
			\label{deploy1}
			\vspace{-2mm}
		\end{minipage}%
	}
	\subfigure[ 6 UAVs, coverage rate = $100\%$,       
	transmission rate = 12.13 Mbps.]
	{
		\begin{minipage}[b]{0.3\linewidth}
			\centering
			\includegraphics[width=2.in,height=1.7in]{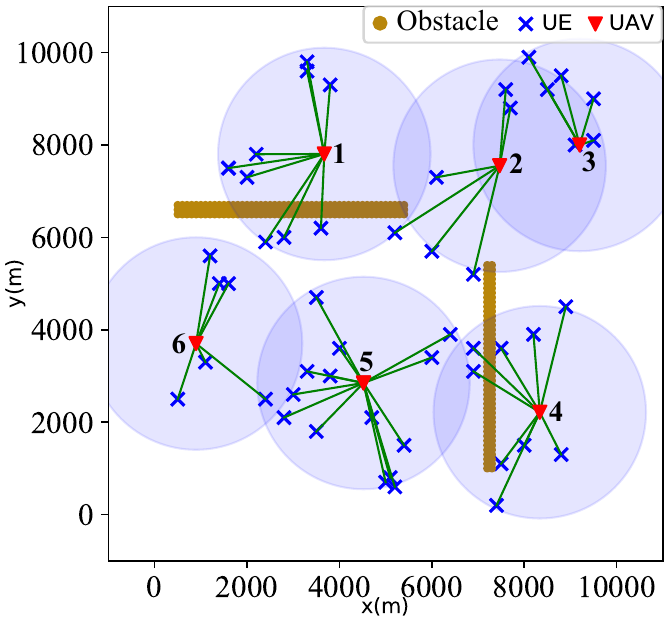}
			\label{deploy2}
			\vspace{-2mm}
		\end{minipage}
	}
	\subfigure[6 UAVs, coverage rate = $100\%$,   transmission rate = 13.98 Mbps.]
	{
		\begin{minipage}[b]{0.3\linewidth}
			\centering
			\includegraphics[width=2.in,height=1.7in]{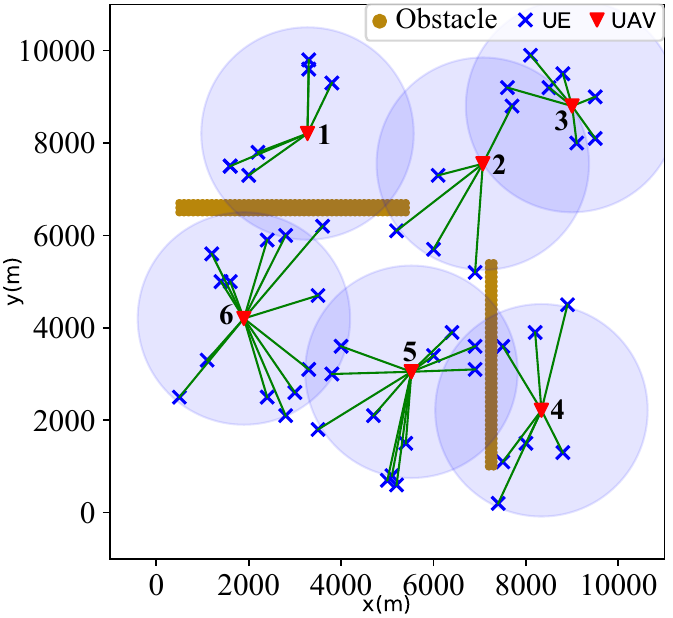}
			\label{deploy4}
			\vspace{-2mm}
		\end{minipage}
	}
	\caption{The results of UAV deployment, power allocation, and user association.}
	\label{figprocess}
	\vspace{-3mm}
\end{figure*}
\begin{algorithm}[t]\small
	\label{algorithm1}
	\textbf{Input:} $\mathcal{K}$, $\varepsilon$, $\gamma_{k}^{th}$, $\mathcal{Z^{\text{ini}}}$, $\mathcal{P^{\text{ini}}}$.\\
	\textbf{Output:} $\mathcal{Z^*}$, $\mathcal{P^*}$, $\mathcal{C^*}$.\\
	\textbf{Initialization:} $N$, $\mathcal{I}$, 
	${\mathcal{Z}}^{(0)}=\mathcal{Z}^{\text{ini}}$, ${\mathcal{P}}^{(0)} = \mathcal{P}^{\text{ini}}$\\
	Calculate $\sum\limits_{i \in \mathcal{I}} \sum\limits_{k \in \mathcal{K}} R_{i,k}^{(0)}$ by \eqref{Rate}.\\
	
	\textbf{Joint Deployment and Power Allocation Optimization:}\\
	\While{$t\leq T$}{
		obtain ${\mathcal{C}}^{(t+1)}$ by coalition game in Section \ref{sec_game},\\
		update ${\boldsymbol{z}}_i^{(t+1)}, \forall i \in \mathcal{I}$  by \eqref{updateu} with fixed ${\mathcal{P}}^{(t)}$,  ${\mathcal{Z}}^{(t+1)} \leftarrow \{{\boldsymbol{z}}_1^{(t+1)},\dots,{\boldsymbol{z}}_{N}^{(t+1)}\}$;\\
		update ${p}_{i,k}^{(t+1)},\forall i \in \mathcal{I},\forall k \in \mathcal{K}$ by \eqref{updatep} with fixed ${\mathcal{Z}}^{(t+1)}$, ${\mathcal{P}}^{(t+1)} \leftarrow \{{p}_{1,1}^{(t+1)},\dots,{p}_{{N},{K}}^{(t+1)}\}$;\\
		calculate $\sum\limits_{i \in \mathcal{I}} \sum\limits_{k \in \mathcal{K}} R_{i,k}^{(t+1)}$ by \eqref{Rate},\\
		$t \leftarrow t+1$.\\
	}
	
	Output $\mathcal{Z}^*={\mathcal{Z}}^{(t)},\mathcal{P}^*={\mathcal{P}}^{(t)},\mathcal{C}^*={\mathcal{C}^{(t)}}$.
	\caption{Virtual Force Approach for UAV Deployment and Power Allocation}
\end{algorithm}

After obtaining the initial deployment for coverage requirement, the user association, deployment, and transmit power of UAVs are jointly optimized to maximize the transmission rate. Consider ${\mathcal{Z}}^{(0)}=\mathcal{Z}^{\text{ini}}$,  $\mathcal{P}^{(0)}=\mathcal{P}^{\text{ini}}$, and the step $t=0$.
At $t$-th iteration, obtain the user association $\mathcal{C}^{(t)}$ by coalition game in Section \ref{sec_game} and given transmit power set $\mathcal{P}^{(t)}$ UAV $i$ updates its location with 
	\begin{equation}
		\label{updateu}
		{\boldsymbol{z}}_i^{(t+1)} = {\boldsymbol{z}}_i^{(t)} + \overrightarrow{{v_i}({\boldsymbol{z}}_i^{(t)},\{{p}_{i,k}^{(t)}\})},
	\end{equation}
and $\mathcal{Z}^{(t+1)}=\{{\boldsymbol{z}}^{(t+1)}_1,\cdots,{\boldsymbol{z}}^{(t+1)}_{N}\}$. The power allocation is updated with ${\mathcal{Z}}^{(t+1)}$ as:
	\begin{equation}
		\label{updatep}
		{p_{i,k}}^{(t+1)}={p_{i,k}}^{(t)}+||\overrightarrow{{v_i}({\boldsymbol{z}_i}^{(t+1)},\{{p_{i,k}}^{(t)}\})}||.
	\end{equation}
\par Finally, the optimal user association $\mathcal{C}^*$, deployment $\mathcal{Z}^*$, and power allocation $\mathcal{P^*}$  are obtained when step exceeds the maxmimum iteration limit $T$.
The details of the aforementioned procedure are summarized in Algorithm \ref{algorithm1}.
\vspace{-3mm}
\section{Simulation Results}
\label{sec_simulation}
\par In this section, simulations are conducted to validate the effectiveness of the proposed algorithm. UEs are distributed within a 10 km x 10 km area. In all simulations, we set the flying height of UAVs is $h=200$ m. Each UAV has the same maximum transmit power $P_1=\dots=P_N = 38$ dBm. The bandwidth of each UAV is set to be $B_1=\dots=B_N=2$MHz. The minimum SNR threshold at the UE is set to be $\gamma_1^{th}=\dots=\gamma^{th}_K = -5$ dB. The carrier frequency is $f=1.4$ GHz. The noise power is $\sigma^2 = -130$ dBm. The empirical found parameters in forest dedicated channel are $A=0.25$, $C=0.39$, $D=0.25$, $E=0$, and $G=0.05$, respectively. The path loss exponent is $\alpha = 3.5$, the standard deviation of shadow fading is $\sigma_x=6$ dB and the reference distance is $d_0 = 1$m. The attractive factor $\lambda_a$ is 1000 and the repulsive force factor $\lambda_r$ is 300. The predefined convergence threshold $\varepsilon$ is $10^{-4}$. 

\begin{figure}[!hbt]
	\centering
	\includegraphics[width=2.5in,height=1.23in]{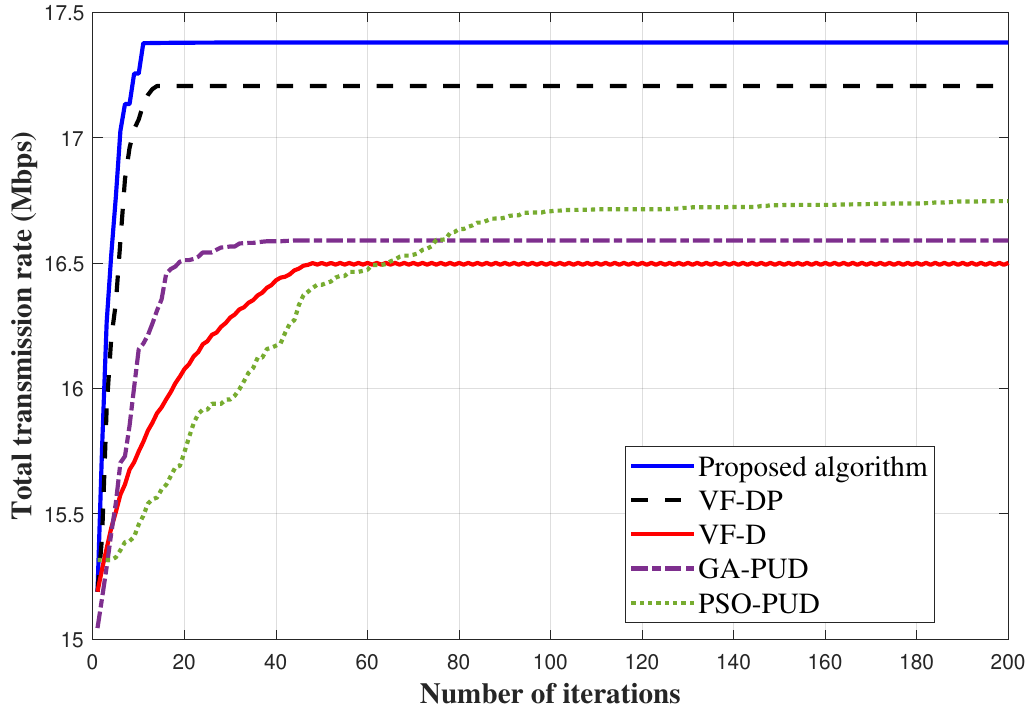}
	\caption{The convergence behavior of the proposed algorithm and its benchmarks.}
	\label{convergence}
	\vspace{-3mm}
\end{figure}
\begin{figure}[!hbt]
	\centering
	\includegraphics[width=2.5in,height=1.23in]{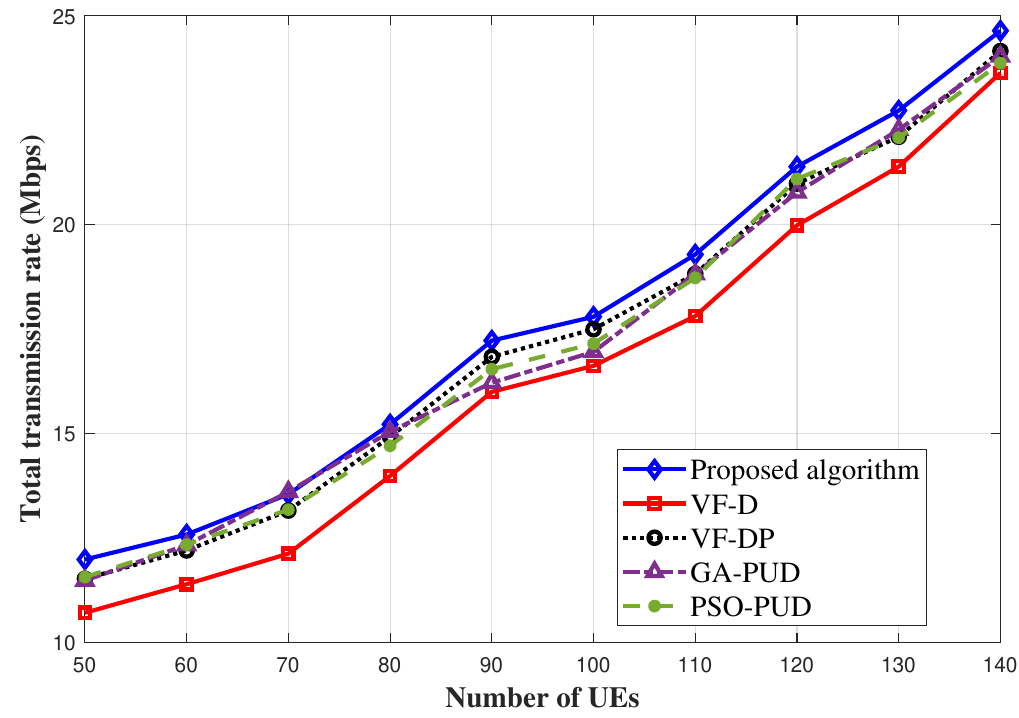}
	\caption{Total transmission rate versus the number of UEs.}
	\label{rate}
	\vspace{-3mm}
\end{figure}
\begin{figure}[!hbt]
	\centering
	\includegraphics[width=2.5in,height=1.23in]{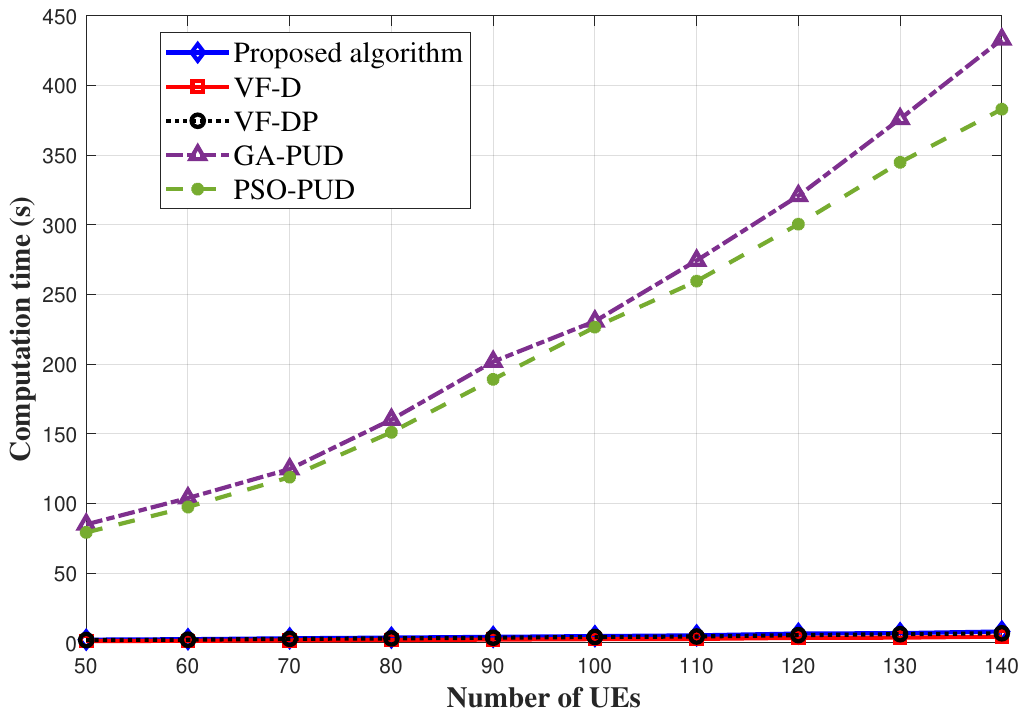}
	\caption{Computation time versus the number of UEs.}
	\label{time}
	\vspace{-3mm}
\end{figure}
Fig. \ref{figprocess} shows the process of UAV deployment, power allocation, and user association by our proposed algorithm. There are 50 UEs, and the number of UAVs increases from Fig. \ref{figprocess}(a) to Fig. \ref{figprocess}(b) to ensure coverage for all UEs. It is observed that 6 UAVs are sufficient for communication. 
Once the number of UAVs is determined, the virtual force and coalition game approach work to further increase the transmission rate by our proposed algorithm.
The initial deployment shown in Fig. \ref{figprocess}(b) disregards the presence of obstacles in the area, which can lead to non-line-of-sight (NLoS) conditions. In contrast, the proposed approach takes into account the influence of obstacles and jointly optimizes the deployment, power allocation, and user association in Fig.  \ref{figprocess}(c). As a result, the total transmission rate increases from $12.13$ Mbps to $13.98$ Mbps.

\par Fig. \ref{convergence} compares the convergence of the proposed algorithm with four benchmarks: the virtual force based joint deployment and power allocation (VF-PD), the virtual force based deployment-only algorithm (VF-D) \cite{8796414}, the GA-based joint deployment, power allocation, and user association algorithm (GA-PUD), and the PSO-based joint deployment, power allocation, and user association algorithm (PSO-PUD)\cite{abdel2021pso}. All the algorithms achieve the maximum transmission rate with multiple step sizes. However, GA-PUD and PSO-PUD exhibit a significantly slower convergence rate, exceeding $70\%$ compared to other methods, primarily due to their high computational complexity. This demonstrates that the virtual force approach outperforms heuristic algorithms in terms of computational efficiency.
\par Fig. \ref{rate} compares the transmission rate and computation time for different numbers of UEs. Our proposed algorithm achieves a comparable transmission rate to GA-PUD and PSO-PUD when considering joint deployment, power allocation, and user allocation optimization, and outperforms VF-D and VF-DP by $7\%$. Additionally, Fig. \ref{time} illustrates that our proposed algorithm achieves a significant reduction in computation time, requiring only $5.6\%$ of the computation time compared to the benchmark schemes.
Therefore, our proposed algorithm offers improved cost-efficiency and better performance.
\vspace{-2mm}
\section{Conclusion}
\par In this paper, we proposed a coalition game-based virtual force algorithm to optimize UAV deployment, power allocation, and user association for demand and enviornmrnt-aware emergency communication. Simulation results demonstrated that the computation time of our proposed algorithm was only $5.6\%$ of the traditional heuristic algorithms such as GA-PUD and PSO-PUD, which is more applicable for disaster scenarios.
\vspace{-5mm}
\bibliographystyle{IEEEtran}
\bibliography{references.bib}
\end{document}